\documentclass[10pt,twocolumn,letterpaper]{article}

\usepackage[pagenumbers]{wacv} 

\usepackage{xcolor}

\definecolor{wacvblue}{rgb}{0.21,0.49,0.74}
\usepackage[pagebackref,breaklinks,colorlinks,allcolors=wacvblue]{hyperref}

\def\wacvPaperID{546} 
\def\confName{WACV}
\def\confYear{2026}

\title{MEGA-PCC: A Mamba-based Efficient Approach for Joint Geometry and Attribute Point Cloud Compression}

\author{
    Kai-Hsiang Hsieh$^1$,  Monyneath Yim$^1$, Wen-Hsiao Peng$^2$, Jui-Chiu Chiang$^{1}$ \\
    $^1$National Chung Cheng University, Taiwan \\
    $^2$National Yang Ming Chiao Tung University, Taiwan \\
    \texttt{kaihsiang@alum.ccu.edu.tw, yimmonyneath@gmail.com,} \\
\texttt{wpeng@cs.nycu.edu.tw, rachel@ccu.edu.tw}
}

\begin{document}
\maketitle
\begin{abstract}
Joint compression of point cloud geometry and attributes is essential for efficient 3D data representation. Existing methods often rely on post-hoc recoloring procedures and manually tuned bitrate allocation between geometry and attribute bitstreams in inference, which hinders end-to-end optimization and increases system complexity. To overcome these limitations, we propose MEGA-PCC, a fully end-to-end, learning-based framework featuring two specialized models for joint compression. The main compression model employs a shared encoder that encodes both geometry and attribute information into a unified latent representation, followed by dual decoders that sequentially reconstruct geometry and then attributes. Complementing this, the Mamba-based Entropy Model (MEM) enhances entropy coding by capturing spatial and channel-wise correlations to improve probability estimation. Both models are built on the Mamba architecture to effectively model long-range dependencies and rich contextual features. By eliminating the need for recoloring and heuristic bitrate tuning, MEGA-PCC enables data-driven bitrate allocation during training and simplifies the overall pipeline. Extensive experiments demonstrate that MEGA-PCC achieves superior rate-distortion performance and runtime efficiency compared to both traditional and learning-based baselines, offering a powerful solution for AI-driven point cloud compression.
\end{abstract}    
\begin{figure}
    \centering
    \includegraphics[width=1\linewidth]{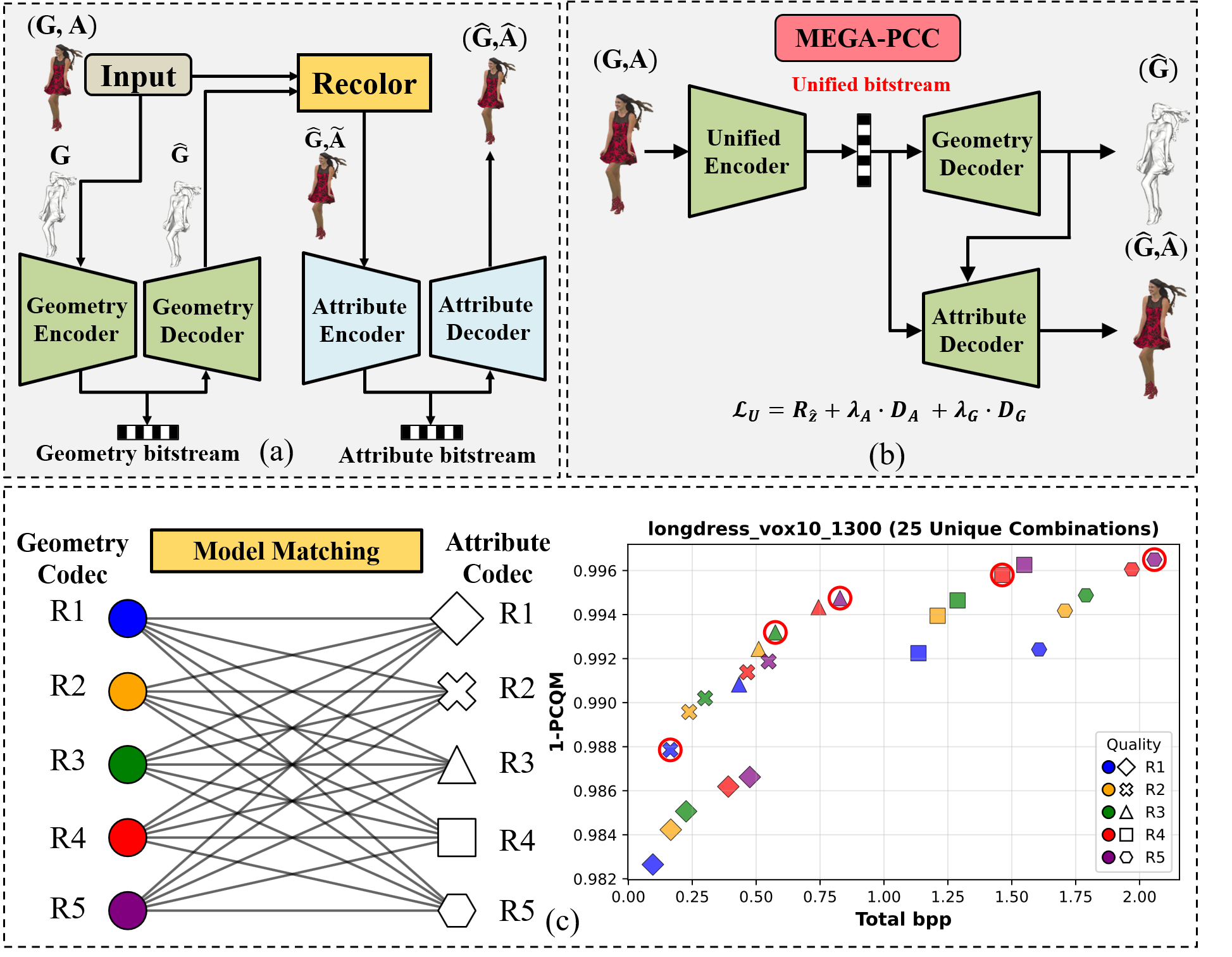}
    \vspace{-7mm}
    \caption{(a) Existing point cloud joint compression methods rely on recoloring to align geometry and attributes. However, manual bit allocation leads to suboptimal reconstruction. (b) The proposed MEGA-PCC uses a shared latent space and loss-based bit allocation, enabling end-to-end compression without recoloring or post-hoc model matching. (c) Traditional model matching involves an exhaustive search of the optimal pairing (e.g., pairing 5 rate points from PCGCv2~\cite{refer8} with ANF-PCAC~ \cite{refer22}) and selects top results based on PCQM scores, incurring high computational cost.}
    \vspace{-5mm}
    \label{fig:Teaser}
\end{figure}



\vspace{-2mm}
\section{Introduction}
\label{sec:intro}

The growing demand for immersive technologies such as the metaverse, virtual and augmented reality (VR/AR), autonomous driving, and telepresence has accelerated the need for efficient 3D data representation. Among various formats, 3D point clouds have become a dominant medium due to their ability to capture detailed geometric structures and rich spatial information. Typically, a point cloud may consist of millions of unordered points, each described by geometric coordinates and attributes like color, surface normals, or reflectance. However, the sheer volume of information involved and irregular structure of point clouds lead to significant challenges in processing and compression, prompting active research into efficient point cloud compression (PCC) techniques.

Recognizing these challenges, the Moving Picture Experts Group (MPEG) introduced two major compression standards: Video-based PCC (V-PCC) and Geometry-based PCC (G-PCC). V-PCC \cite{refer1} projects 3D point clouds onto 2D patches and leverages mature video codecs like H.265/HEVC \cite{refer3}, making it well-suited for dynamic point clouds. In contrast, G-PCC \cite{refer2} directly compresses 3D geometry using octree-based partitioning and encodes attributes using tools such as Region Adaptive Hierarchical Transform (RAHT) \cite{refer4}, enabling efficient compression of both static and dynamic content. While these standards laid the foundation for practical point cloud compression, they rely on manually designed pipelines with limited flexibility and task-specific design. In recent years, learning-based PCC has emerged as a powerful alternative, offering data-driven solutions that can learn compact latent representations, adapt to diverse point cloud structures, and optimize rate-distortion performance in an end-to-end fashion.

\begin{figure}
    \centering
    \includegraphics[width=1\linewidth]{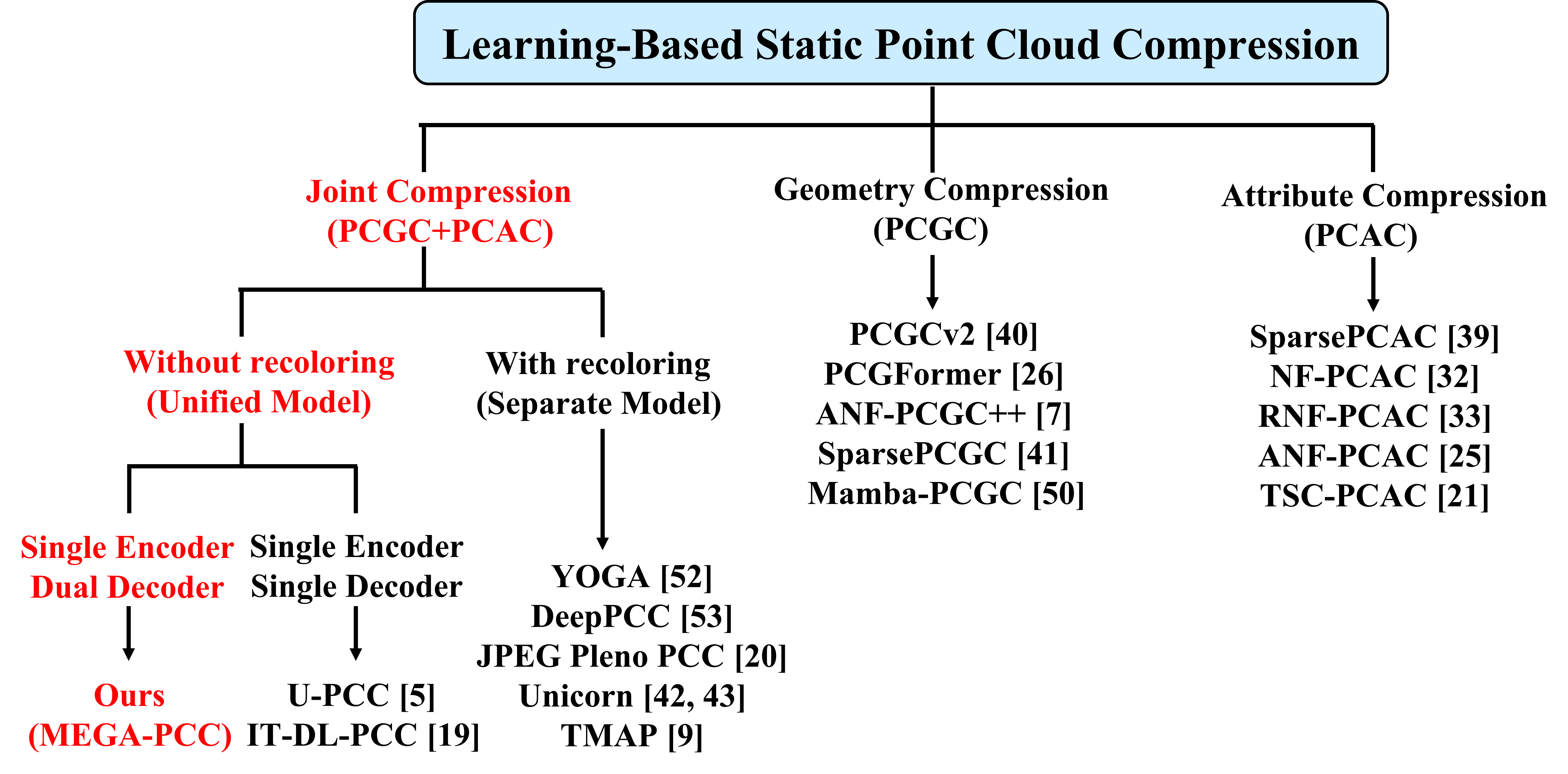}
    \vspace{-7mm}
    \caption{Taxonomy of point cloud compression methods.}
    \vspace{-5mm}
    \label{fig:Taxonomy}
\end{figure}

Early research in learned point cloud compression predominantly treated geometric information and attribute data as independent optimization problems, resulting in separate compression pipelines for each modality. As the field progressed, joint compression approaches emerged as a promising research direction, aiming to exploit the correlation between geometry and attributes for improved efficiency. Figure ~\ref{fig:Taxonomy} illustrates the evolution of point cloud compression techniques, categorizing them into three main branches: geometry compression, attribute compression, and joint compression. Initial attempts at joint compression, such as U-PCC~\cite{refer24} and IT-DL-PCC \cite{refer25} represent point clouds as multi-channel 3D voxel grids, with separate channels for occupancy and colors. Although conceptually unified, these voxel-based methods often suffer from suboptimal rate-distortion performance. A more common direction, illustrated in Figure~\ref{fig:Teaser}(a), involves coordinating separate models for geometry and attribute compression, typically connected through an intermediate process like recoloring. This design surpasses G-PCC in performance and demonstrates enhanced coding efficiency. However, these schemes still suffer from two critical drawbacks:

First, optimal bit allocation between geometry and attributes remains a key challenge due to their strong interdependence. Poor allocation can significantly impair reconstruction quality. For instance, allocating too few bits to geometry leads to inaccurate point positions, which in turn distorts attributes during the geometry-dependent recoloring process. Conversely, under-allocating bits to attributes directly degrades visual quality. To mitigate this, many joint compression methods resort to an exhaustive model matching process, evaluating multiple combinations of geometry and attribute compression settings at a fixed bitrate to select the best-performing pair, as illustrated in Figure~\ref{fig:Teaser}(c). While effective, this strategy is computationally intensive and limits deployment in real-time or large-scale scenarios. 

Second, the recoloring step introduces architectural constraints that break the continuity of the compression pipeline. This intermediate operation, which maps attributes onto compressed geometry, becomes a bottleneck that disrupts seamless integration. Errors introduced in geometry compression propagate through recoloring, making attribute quality tightly dependent on geometric fidelity.  Additionally, the sequential dependency imposed by recoloring—requiring geometry to be processed before attribute compression, slows down the overall pipeline and blocks joint learning, further limiting efficiency and adaptability.


These limitations underscore the urgent need for a more unified and efficient compression framework that can jointly handle geometry and attribute data, support end-to-end optimization, and perform adaptive bit allocation without relying on post-hoc tuning or computationally expensive model matching. To address these challenges, we propose MEGA-PCC - a unified, end-to-end point cloud compression framework that simultaneously compresses geometry and attributes using a shared latent representation, as illustrated in Figure~\ref{fig:Teaser}(b). MEGA-PCC features a single-encoder, dual-decoder architecture, where geometry and attributes are encoded together into a shared latent space that captures their mutual dependencies. The shared latent not only eliminates the need for separate encoding paths but also removes the need for recoloring, thus resolving the error propagation and sequential bottlenecks of prior joint compression pipelines. MEGA-PCC enhances its encoder/decoder by integrating 3D sparse convolutions with State Space Models (SSMs) based on the Mamba architecture. This hybrid design leverages Mamba's hardware-aware, selective scan mechanism to achieve linear time complexity and sub-quadratic memory usage, allowing the model to efficiently capture both local geometric patterns and global spatial dependencies. Furthermore, we introduce a Mamba-based entropy model to better capture channel and spatial correlations during entropy coding, which further improves rate-distortion efficiency.

Importantly, MEGA-PCC supports end-to-end training with integrated bit allocation, bypassing the need for model matching by enabling the encoder-decoder system to dynamically learn how to distribute bits between geometry and attributes. During inference, geometry is decoded first and used to guide the attribute decoder, ensuring accurate attribute reconstruction while preserving a lightweight architecture suitable for efficient deployment.

The main contributions of this work are as follows:

\begin{itemize}
    \item MEGA-PCC introduces a novel single-encoder, dual-decoder design that jointly compresses geometry and attributes, avoiding both the  recoloring step and the need for coordinating separate models.
    
\end{itemize}

\begin{itemize}
    \item MEGA-PCC leverages Mamba architecture to enhance both the encoder/decoder and the entropy model, improving the modeling of long-range dependencies and boosting compression performance.
\end{itemize}

\begin{itemize}
    \item MEGA-PCC supports fully end-to-end training and learns to allocate bits between geometry and attributes without requiring exhaustive model matching, thus achieving efficient and fast compression.
\end{itemize}

\section{Related Work}
\label{sec:formatting}
\vspace{-1mm}


\subsection{Point Cloud Processing}


Early point-based methods like PointNet \cite{refer5} and PointNet++ \cite{refer6} enabled deep learning on raw, unordered point clouds, but struggled to capture both local context and global structure. This led to attention-based methods such as Point Transformer \cite{refer12,refer49,refer50}, which introduced tailored self-attention for improved local feature aggregation. Building on this, Point-BERT \cite{refer47} and Point-MAE \cite{refer48} applied standard Transformers for self-supervised pretraining on point cloud data. Despite performance gains, these models face quadratic costs due to global attention. Alternatively, voxel-based methods convert point clouds into structured 3D grids \cite{refer56, refer57, refer58, refer59, refer60}, enabling efficient processing via sparse convolutions or voxel-based Transformers. However, they suffer from voxelization limitations, such as high memory usage and limited receptive fields, which hinder dense point cloud processing.

To overcome the limitations of Transformer-based models, recent work has explored State Space Models (SSMs) \cite{refer10, refer45}, which model long-range dependencies with linear complexity. When applied to point clouds \cite{refer15, refer16, refer61}, SSMs require serializing 3D data into sequences—a critical step for maintaining spatial coherence and fully leveraging sequential modeling capabilities. The effectiveness of SSMs like Mamba hinges on this serialization process, as the scan order significantly affects spatial continuity and local feature preservation. To address this, Point Cloud Mamba \cite{refer15} uses grid sampling to assign spatial codes that keep neighboring points close in the sequence. Mamba3D \cite{refer16} further improves spatial feature capture by introducing a bidirectional scan strategy that traverses both spatial and channel dimensions, reducing sensitivity to serialization order.

\begin{figure*}[t]
  \centering
\includegraphics[width=0.95\textwidth, keepaspectratio]{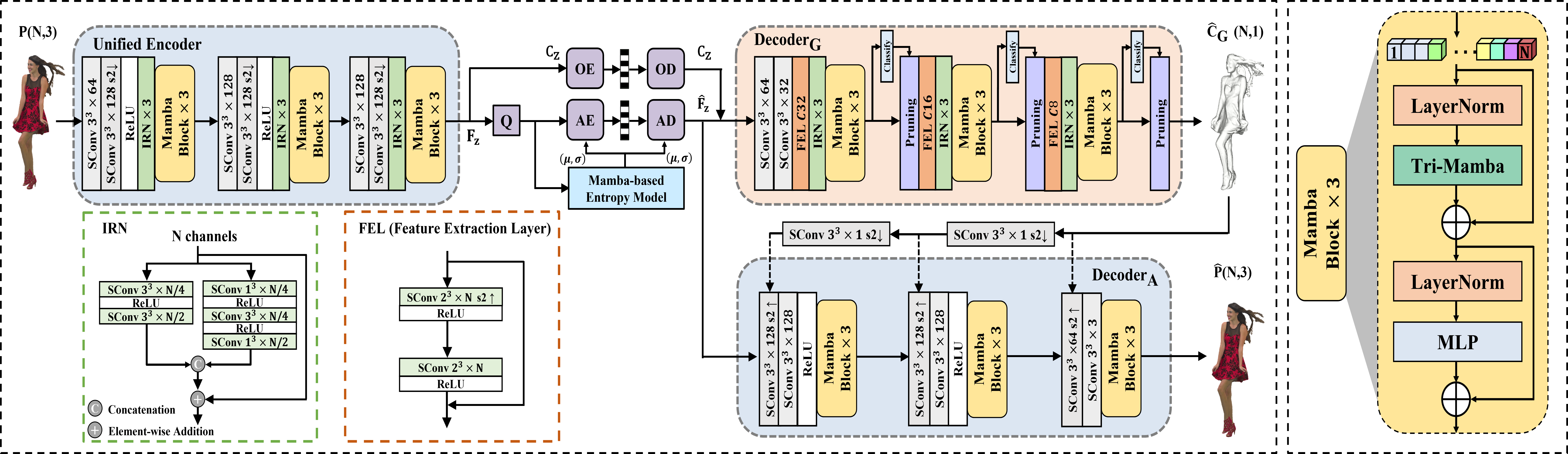}
\vspace{-2.5mm}
    \caption{The unified encoder jointly processes geometry and attributes using multi-layer sparse convolutions and Mamba blocks to capture both local and global dependencies, producing a shared latent representation. During decoding, geometry is reconstructed first, and the decoded coordinates guide the attribute decoder, enabling efficient joint compression.}
    \label{fig:MEGA_framework}
     \vspace{-4.5mm} 
\end{figure*}
\subsection{Point Cloud Compression}
\textbf{Geometry Compression.} Learning-based point cloud compression has evolved along several primary methodological branches. We begin by briefly reviewing approaches to point cloud geometry compression.  Point-based approaches \cite{refer51, refer52}, operate directly on individual points to preserve fine spatial details. Octree-based approaches \cite{refer53, refer54, refer55} recursively partition space into tree structures, enabling efficient geometry representation. By modeling occupancy based on ancestor and sibling nodes, they are well-suited for large-scale, sparse data like LiDAR, Voxel-based methods \cite{refer8, refer9, refer11, refer17, refer18, refer19, refer20, refer21, refer22, refer23, refer24, refer25, refer26, refer27, refer28, refer29, refer31, refer32, refer33} convert irregular point clouds into structured voxel grids, simplifying spatial representation and offering effective compression for more uniformly distributed data. Early works like PCGCv2 \cite{refer8} use sparse tensors to balance fidelity and efficiency, while PCGFormer \cite{refer9} introduces KNN-based self-attention to overcome fixed receptive fields. Recent advances include ANFPCGC++ \cite{refer17}, integrating augmented normalizing flows with transformer-based context modeling for more accurate probabilistic latent feature representation; SparsePCGC \cite{refer18}, which exploits multi-scale conditional coding with hierarchical priors, using lower-scale information to guide higher-scale reconstruction and boost compression efficiency; and Mamba-PCGC \cite{refer11}, which achieves linear computational complexity while maintaining comprehensive receptive fields to facilitate effective latent feature extraction. \cite{refer62, refer63} introduce a variable rate approach that employs a single model to achieve adaptive rate control.

\noindent\textbf{Attribute Compression.} Beyond geometric compression, efficient point cloud attribute compression (PCAC) is essential for high-fidelity 3D reconstruction and photorealistic rendering. SparsePCAC \cite{refer19} builds on the PCGCv2 framework \cite{refer8} by incorporating an autoregressive context model with a hyperprior, improving entropy estimation and compression accuracy, at a cost of increased decoding complexity. Moving beyond traditional VAE-based approaches, NF-PCAC \cite{refer20, refer21} leverages normalizing flows to enhance attribute reconstruction. ANF-PCAC \cite{refer22} further advances the field by using augmented normalizing flow with additional conditional variables, enabling more expressive modeling of attribute dependencies and yielding better reconstruction quality. Most recently, TSC-PCAC \cite{refer23} integrates Transformer modules and sparse convolutions within a VAE framework, using a channel context module to effectively balance compression performance and computational cost.

\vspace{0.1cm} 
\noindent\textbf{Joint Geometry and Attribute Compression.} Early efforts, such as U-PCC \cite{refer24} and IT-DL-PCC \cite{refer25}, pioneer end-to-end joint compression architectures by directly processing four-channel inputs combining geometry and attribute information. While IT-DL-PCC introduces super-resolution modules to improve performance over G-PCC, both methods rely on single-stage training, making it difficult to balance compactness and reconstruction fidelity. As a result, their performances lag behind more recent approaches. Later methods \cite{refer26, refer27, refer28, refer29, refer31, refer32, refer33} adopt sequential pipelines, where geometry is encoded first, followed by recoloring and attribute compression. For example, YOGA \cite{refer26} combines octree-based geometry coding (via G-PCC), neural enhancement, and RAHT-based attribute reconstruction. DeepPCC \cite{refer27} separately encodes geometry and attributes using two networks, incorporating multiscale neighborhood aggregation through sparse convolutions and local attention. Unicorn \cite{refer31, refer32} proposes a multiscale conditional coding framework that supports both lossy and lossless compression, as well as static and dynamic content. TMAP \cite{refer33}, being developed under MPEG, advances this direction as a unified AI-PCC standard. Meanwhile, the JPEG Pleno PCC VM4.1 standard \cite{refer29} integrates learned geometry compression with JPEG AI \cite{refer30} for attributes, establishing a new benchmark in AI-driven point cloud compression. 

Despite progress, most methods still rely on recoloring to connect geometry and attribute pipelines. This decoupled design hinders end-to-end optimization and requires manual geometry-attribute matching to balance rate-distortion trade-offs, increasing complexity and limiting adaptability across content types and bitrates. These issues underscore the need for unified frameworks that jointly model geometry and attributes without intermediate steps, enabling flexible optimization.



\section{Proposed Method}
\vspace{-1mm}
\subsection{Preliminary}
\vspace{-2mm}
State space models \cite{refer10} (SSMs) offer a principled framework for modeling dynamical systems by capturing the relationship between input sequences and outputs through latent hidden states. 

Given an input sequence $\mathbf{x}(t) \in \mathbb{R}^{L}$, the system maintains a hidden state $\mathbf{h}(t) \in \mathbb{R}^{H}$ and generates an output $\mathbf{y}(t) \in \mathbb{R}^{L}$ by linear ordinary differential equations (ODEs):

\begin{equation}
\begin{aligned}
\mathbf{h}'(t) &= \mathbf{A} \mathbf{h}(t) + \mathbf{B} \mathbf{x}(t), \\
\mathbf{y}(t) &= \mathbf{C} \mathbf{h}(t),
\end{aligned}
\end{equation}

where \( \mathbf{A} \in \mathbb{R}^{H \times H} \), \( \mathbf{B} \in \mathbb{R}^{H \times L} \), and \( \mathbf{C} \in \mathbb{R}^{L \times H} \) are learnable parameters defining the dynamics and projection mappings of the system. To apply SSMs in digital systems, the continuous-time formulation must be discretized. A common approach is the Zero-Order Hold (ZOH) method, which approximates the system using a fixed time step 
$\Delta$:
\begin{equation}
\vspace{-2mm}
    \begin{array}{l}
\bar{\mathbf{A}} = \exp(\Delta \mathbf{A}), \\
\bar{\mathbf{B}} = (\Delta \mathbf{A})^{-1} (\exp(\Delta \mathbf{A}) - \mathrm{I}) \cdot \Delta \mathbf{B}.
\end{array}
\vspace{1mm}
\end{equation}
resulting in the discrete-time update rules:
\begin{equation}
\begin{aligned}
\mathbf{h}_t &= \overline{\mathbf{A}} \mathbf{h}_{t-1} + \overline{\mathbf{B}} \mathbf{x}_t, \\
\mathbf{y}_t &= \mathbf{C} \mathbf{h}_t.
\end{aligned}
\vspace{1mm}
\end{equation}

This discrete form enables efficient integration of SSMs into modern deep learning models, offering linear-time complexity while capturing long-range dependencies.

\subsection{Overall framework}

Geometry and attributes in point clouds are inherently interdependent—attributes are defined at specific 3D coordinates, and their accurate representation relies on the underlying spatial structure. This intrinsic coupling motivates unified compression frameworks that model both modalities jointly, rather than in separate stages. To address this, we propose MEGA-PCC, a unified and end-to-end trainable framework that simultaneously compresses geometry and attributes, as illustrated in Figure~\ref{fig:MEGA_framework}. MEGA-PCC extends Mamba-PCGC \cite{refer11}—originally designed for geometry-only compression—by incorporating attribute representation into a shared latent space. The input point cloud is voxelized into a sparse 3D grid, where each point is represented by geometric coordinates $C = {(x_i, y_i, z_i) \mid i \in [0, N-1]}$ and attributes $F = {(R_i, G_i, B_i) \mid i \in [0, N-1]}$ where $N$ is the total number of points. The coordinates and attributes are jointly encoded into a three-channel volumetric input.

The unified encoder combines sparse convolutions for local structure modeling with multi-directional SSMs to capture long-range dependencies. Unlike the unidirectional Mamba block in \cite{refer11}, MEGA-PCC adopts three directional SSM modules, including Forward SSM, Backward SSM, and Channel Flip SSM to model spatial context more comprehensively along different axes.  

During encoding, the input is transformed into a compact latent representation $z$, consisting of a geometric skeleton $C_z$ and a feature tensor $F_z$. The structural skeleton is losslessly encoded using G-PCC to preserve occupancy information, while $F_z$ is quantized and entropy encoded using the proposed Mamba-based Entropy Model (MEM). In decoding, the quantized features $\hat{F}_z$ are first processed by the geometry decoder to reconstruct the point coordinates $\hat{C}_G$, following a coarse-to-fine strategy that includes a Top-$k+1$ selection mechanism \cite{refer34}. Subsequently, the reconstructed geometric coordinates $\hat{C}_G$ guide the attribute decoder to reconstruct the attributes in a coarse-to-fine approach, leveraging the multi-scale geometric information. This tightly coupled decoding process ensures that geometry and attributes are jointly optimized, resulting in coherent and efficient compression.

\subsection{Mamba Block in Encoder and Decoder}

\begin{figure}
    \centering
    \includegraphics[width=1\linewidth]{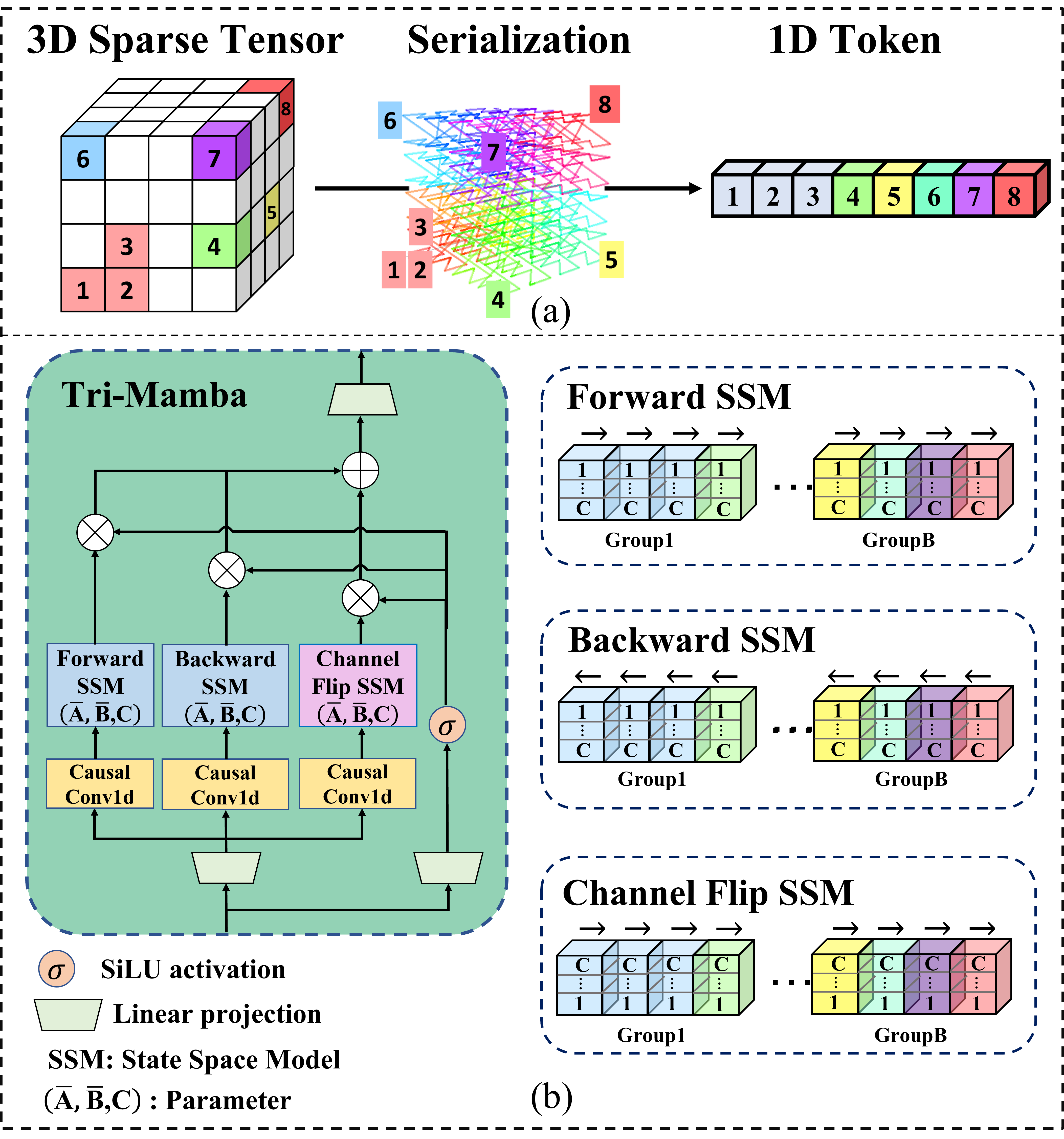}
    \vspace{-7mm}
    \caption{(a) Serialize point clouds into a sequence while preserving spatial proximity relationships between consecutive elements in the sequence (b) Tri-Mamba, which is used in both encoder and decoder for feature extraction, combines forward, backward, and feature channel scanning to comprehensively understand spatial information and leverage channel-wise information to enrich feature representation.}
    \label{fig:mambascan}
    \vspace{-6mm} 
\end{figure}

 To enhance spatial feature modeling in the proposed scheme, we introduce the Mamba block, used throughout the encoder and decoder of MEGA-PCC. Its architecture is illustrated in Figure~\ref{fig:mambascan}. Since Mamba operates on 1D sequences, we serialize the 3D voxel grid using a Morton scan~\cite{refer35} (Figure~\ref{fig:mambascan}(a)). This scan preserves spatial locality during flattening, converting the 3D sparse tensor into a 1D sequence while maintaining the neighborhood structure to the extent possible. The resulting latents are then divided into groups, 
 and processed independently by dedicated Mamba modules, enabling parallel computation across groups while allowing autoregressive modeling within each group. However, relying solely on Morton-based serialization may not fully capture complex spatial dependencies. Inspired by Vision Mamba \cite{refer45} and Mamba3D \cite{refer16}, we propose Tri-Mamba in a Mamba block (Figure~\ref{fig:mambascan}(b)), which integrates three complementary scanning directions: Forward SSM processes natural sequence order, Backward SSM handles reverse order for comprehensive context, and Channel Flip SSM explores inter-channel relationships, thereby reducing pseudo-dependence on order. By aggregating the outputs from all three directions into unified representations, Tri-Mamba effectively captures both fine-grained spatial details and global context, which is critical for accurate joint compression.
 
The Mamba block is deployed across three downsampling stages in the encoder, each stage progressively reducing point cloud resolution. To maintain efficiency, we adopt adaptive group sizing: larger blocks are used at higher-density levels, and smaller blocks are used as the resolution decreases. We denote the group sizes as $\mathbf{M} = [\mathbf{M}_1, \mathbf{M}_2, \mathbf{M}_3]$ for the three stages, respectively.


\subsection{Mamba-based Entropy Model}
\begin{figure}
    \centering
    \includegraphics[width=1\linewidth]{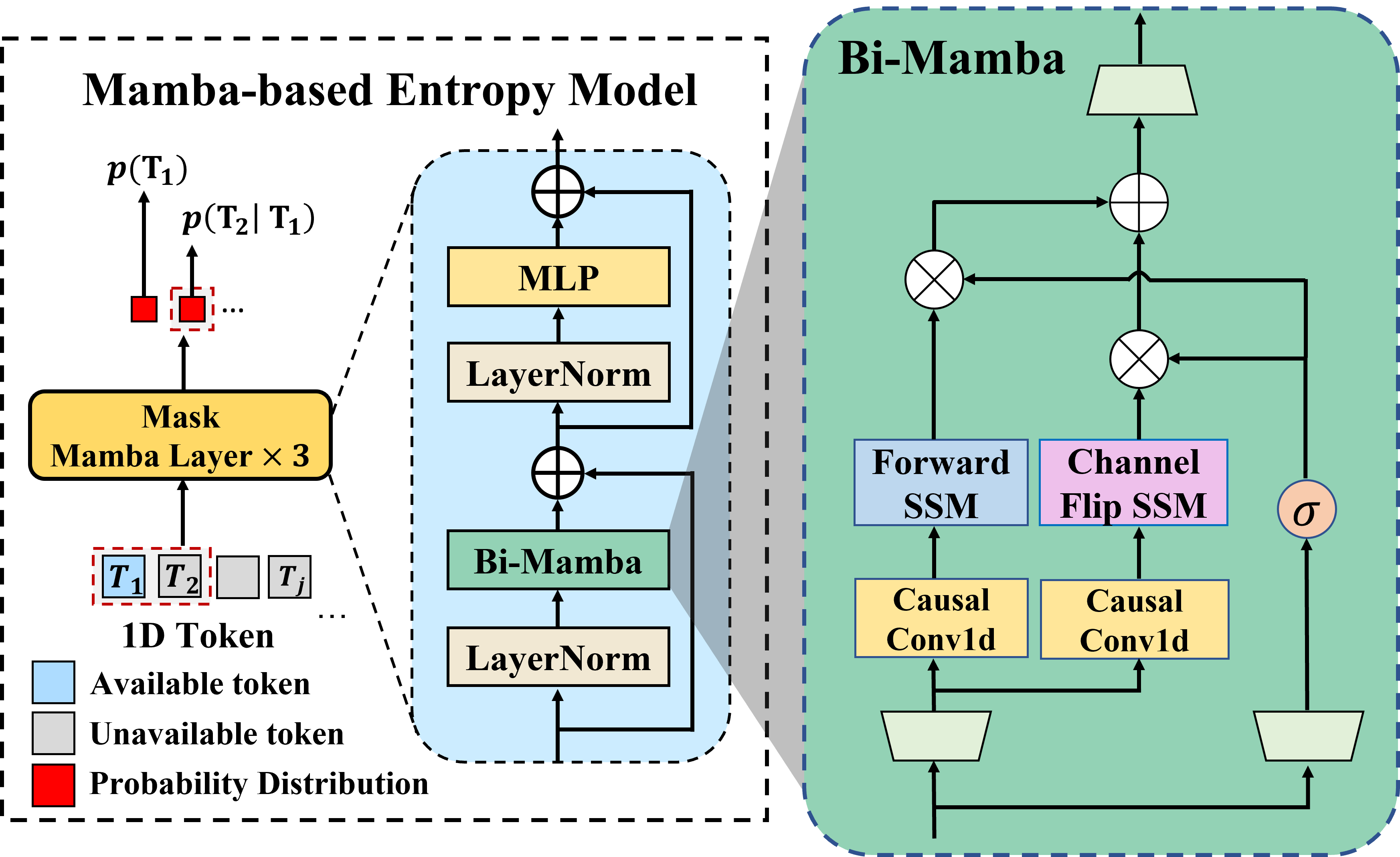}
    \vspace{-5mm}
    \caption{Mamba-based Entropy model.}
    \label{fig:MEM}
    \vspace{-5mm} 
\end{figure}

Accurate entropy modeling in point cloud compression remains a major challenge due to the irregular and sparse structure of point clouds. A common baseline is the factorized entropy model, which assumes independence across spatial positions or feature channels. However, such simplification fails to account for the rich spatial and semantic correlations inherent in point cloud data. Recent approaches, such as \cite{refer17}, improve over the factorized model by incorporating spatial context to capture local dependencies. Nonetheless, they still overlook channel-wise correlations, which are critical for modeling interactions across feature dimensions. To overcome these limitations, we propose the Mamba-based Entropy Model (MEM), shown in Figure~\ref{fig:MEM}. MEM introduces a unified context model that captures both spatial and channel-wise dependencies in the latent features, enhancing the accuracy of probability estimation for entropy coding.

As in our encoder, MEM begins by converting the 3D latent volume into a 1D token sequence using a Morton scan to preserves spatial locality. The sequence is then divided into groups of length $J$. Within each group, tokens are processed causally, enabling autoregressive modeling while supporting parallel computation across groups. The core component of MEM is the Bi-Mamba module, which performs two causal scans. Forward SSM processes the token sequence in a forward direction to model long-range spatial relationships, while Channel Flip SSM scans across feature channels to capture inter-channel correlations for each token. Both scanning operations are causal, ensuring that only past information is used when predicting the current token’s distribution. Their outputs are fused to produce conditional probabilities for entropy coding, as shown in Figure~\ref{fig:MEM}. For the token $T_j$, it probability is computed as:
\vspace{-2mm}
\begin{equation}
p(T_1, T_2, \ldots, T_J) = \prod_{j=1}^{J} p\bigl(T_j \mid T_{<j}\bigr),
\label{eq:mem}
\end{equation}
\vspace{-3mm}

\noindent By jointly modeling spatial and channel-wise dependencies, the Bi-Mamba module enables MEM to capture richer context, leading to more accurate probability estimates and improved point cloud compression. 
\subsection{Training Strategy}
Training a unified encoder to jointly represent geometry and attributes poses significant challenges. In single-stage training, the shared latent representation often fails to balance the reconstruction quality of both geometry and attributes effectively. This imbalance hinders the model’s ability to optimize both components simultaneously, resulting in suboptimal compression performance. To address this, we propose a two-stage training strategy. Considering that attribute encoding is inherently more complex than geometry encoding which primarily involves binary occupancy prediction, the first stage focuses exclusively on optimizing attribute compression. The second stage then jointly trains both geometry and attribute components. In the supplementary (SEC. H), we provide additional results with single-stage training, confirming the superiority of two-stage training.

During the first stage, attribute coding is trained assuming that the ground-truth geometry is available. The loss function, described in \cref{eq:attloss}, measures the distortion $D_{A}$ in the YUV color space using the mean squared error (MSE) between the original and the reconstructed point cloud attributes:
\begin{equation}
    \mathcal{L}_{A}=R_{\hat{Z}}+\lambda_{A} \cdot D_{A}.
    \label{eq:attloss}
\end{equation}
In the second stage,  joint training optimizes both geometry and attribute compression simultaneously, balancing their respective distortions using
\cref{eq:totalloss}, where $D_{G}$ is the binary cross-entropy loss for geometry reconstruction. As in many learned compression systems, $\lambda_{A}$ and $\lambda_{G}$ are chosen empirically. Identifying principled strategies for selecting these hyperparameters is part of our future work.

\begin{equation}
\mathcal{L}_{U}=R_{\hat{Z}}+\lambda_{A} \cdot D_{A}+\lambda_{G} \cdot D_{G} 
\label{eq:totalloss}
\end{equation}

To mitigate early training instability and resource contention, we initialize the unified encoder and attribute decoder with the weights from the first stage, while the geometry decoder is randomly initialized. For the first 20 epochs, the attribute decoder is conditioned on ground-truth geometry to prevent inaccuracies in geometry from degrading attribute learning. After the geometry decoder achieves sufficient reconstruction accuracy, its output is used by the attribute decoder for attribute reconstruction.

At this stage, geometry is no longer assumed to be lossless as in the first stage, making attribute compression and distortion computation more complex due to potential misalignment between the reconstructed and original point clouds. To address this, we adopt a bidirectional mean squared error metric, which evaluates distortion in both forward and backward directions. This mitigates bias from asymmetric point matching and provides a more reliable assessment of attribute distortion. 


\section{Experiments}

\begin{table*}[h]
        \centering
        \vspace{-5mm}
        \scriptsize
        \captionof{table}{BD-Rate (\%) comparison for geometry and attribute distortion relative to G-PCCv23}
        \vspace{-2.5mm}
        \label{table:all}
        \setlength{\tabcolsep}{3pt}
        \renewcommand{\arraystretch}{1}
        
        \resizebox{\textwidth}{!}{%
        \begin{tabular}{c|ccc|ccc|ccc|ccc|ccc|ccc}
        \toprule
\multicolumn{1}{c|}{\textbf{\textbf{Sequence}}} & \multicolumn{3}{c|}{\textbf{V-PCCv22 \cite{refer1}}} & \multicolumn{3}{c|}{\textbf{YOGA \cite{refer26}}} & \multicolumn{3}{c|}{\textbf{DeepPCC \cite{refer27}}} & \multicolumn{3}{c|}{\textbf{JPEG Pleno VM \cite{refer29}}} & \multicolumn{3}{c|}{\textbf{Unicorn \cite{refer32}}} &
\multicolumn{3}{c}{\textbf{Proposed}}
\\ 
 & D1 & Y & 1-PCQM  & D1  & Y  & 1-PCQM   & D1  & Y  & 1-PCQM   & D1  & Y  & 1-PCQM  & D1  & Y  & 1-PCQM   & D1  & Y  & 1-PCQM  \\ 
\midrule
longdress & -66.8 & -64.7 & -50.1 & -86.1 & -48.5 & -32.4 & -83.7  & -49.8 & -47.5 & - & - & - & - & - & -63.0 & -79.2 & -41.3 & -47.3 \\ 

loot & -74.4 & -72.2 & -64.9 & -87.0  & -55.9 & -53.2 & -84.9 & -51.0 & -55.4 & -  & - & - & - & - & -77.7 & -81.5  &-54.5 & -58.3  \\

redandblack & -66.1 & -64.7 & -52.2 & -84.9  & -55.1 & -48.5 & -81.8 & -53.2 & -54.3 & -  & - & - & - & - & -65.9 & -74.1  & -54.4 & -52.6  \\ 

soldier & -64.0 & -60.5 & -53.1 & -85.3  & -53.2 & -46.7 & -81.8 & -54.5 & -61.1 & -62.9  & -51.5 & -42.3 & - & - & -74.2 & -78.2  & -52.8 & -58.6 \\ 

basketball\_player & -85.3 & -66.2 & -49.7 & -60.0 & 38.3 & 68.3 & -92.9 & -17.5 & 11.5 & -  & - & - & - & - & - & -85.5 & -31.7 & -40.7  \\ 

dancer & -83.9 & -66.5 & -46.4 & -55.5 & 47.0 & 75.8 & -91.8  & -6.2 & 12.7 & - & - & - & - & - & - & -84.0 & -28.7 & -40.2  \\ 
\midrule

Average & -73.4 & -65.8 & -52.7 & -76.5& -12.9 & -6.1 & -86.2  & -38.7 & -32.4 & -62.9 & -51.5 & -42.3 & - & - & -70.2 & -80.4 & -43.9 & -49.6 \\
    \bottomrule
    \end{tabular}}
    \vspace{-1mm}
    \label{table:BDBR}
    \end{table*}
\begin{figure*}
    \centering
    \includegraphics[width=1\linewidth]{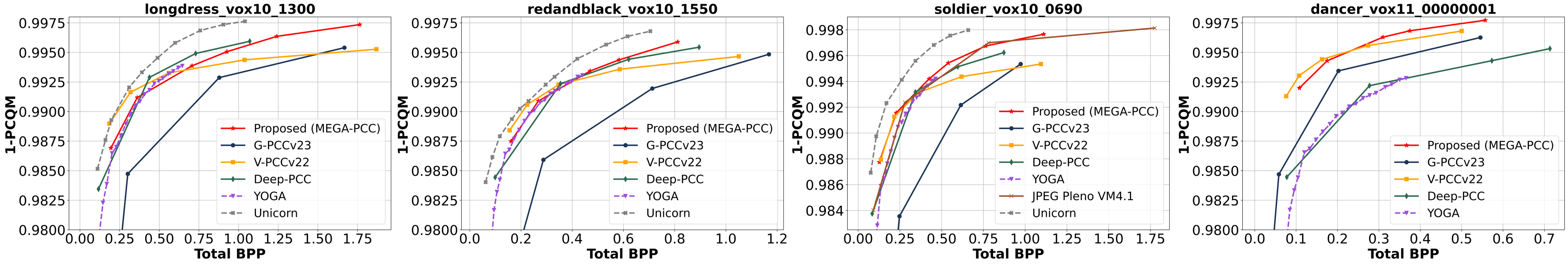}
    \vspace{-7mm}
    \caption{R-D performance of the proposed scheme in terms of 1-PCQM.}
    \vspace{-3mm}
    \label{fig:RDcurve}
\end{figure*}


\subsection{Experimental  Setup}

\textbf{Datasets.} We train our model on two datasets: ScanNet \cite{refer37}, which contains over 1,500 richly annotated indoor 3D scenes, and RWTT \cite{refer38}, known for its detailed color and texture information. To accommodate GPU memory constraints, the original point clouds are partitioned into non-overlapping cubes of size 7 bits, from which 15,000 cubes are sampled for training. For evaluation, we follow the G-PCC Common Test Conditions \cite{refer43} and feed the entire point cloud without partitioning. Testing is conducted on two benchmark datasets. 8i Voxelized Full Bodies (8iVFB) \cite{refer39}, includes four 10-bit sequences: \textit{longdress}, \textit{loot}, \textit{redandblack}, and \textit{soldier}. Owlii dynamic human mesh (Owlii) \cite{refer40} contains two 11-bit sequences: \textit{basketball\_player} and \textit{dancer}.

\noindent\textbf{Training Setting.} MEGA-PCC is implemented using PyTorch and MinkowskiEngine \cite{refer36} and all experiments are conducted on a workstation equipped with an Intel i9-12900K CPU and an NVIDIA RTX 3090 GPU.
Training is performed for up to 60 epochs with random weight initialization. The initial learning rate is set to $4 \times 10^{-5}$ and decays by half every 20 epochs, reaching $5 \times 10^{-6}$ by the end. We use the Adam optimizer with $\beta_1 = 0.9$, $\beta_2 = 0.999$, and a batch size of 8. For the attribute-only training phase (see \cref{eq:attloss}), $\lambda_A$ is set to 0.05. In joint training (see \cref{eq:totalloss}), $\lambda_A$ is set to 0.03, 0.0195, 0.013975, 0.0085, 0.0032, and 0.00135, while $\lambda_G$ is set to 7.2, 1.625, 0.7525, 0.75, 0.36, and 0.225 for training the corresponding models. In the Mamba blocks, we use  \(\mathbf{M} = [1024, 512, 256]$ across three scales. The group size $J$ in the Mamba-based entropy model is fixed at 16.
{The overall training time requires approximately 60 hours.
\vspace{0.1cm} 

\noindent\textbf{Evaluation Metric.}
To comprehensively evaluate point cloud compression performance, we adopt both distortion-based and perceptual quality metrics. D1-PSNR and D2-PSNR are used to assess geometric reconstruction accuracy, while Y-PSNR and YUV-PSNR evaluate color fidelity. However, these conventional metrics may not fully reflect perceptual quality in the context of joint geometry and attribute compression. Hence, we also include PCQM \cite{refer41}, which is recommended by the JPEG Pleno standard for its better alignment with human visual perception. To further complement this evaluation, we also report GraphSIM \cite{refer42}, which uses a radius-based neighborhood to measure local consistency, and PQI \cite{refer44}, which jointly quantifies geometric and visual distortions for a holistic assessment. 

\begin{table*}[h]
    \centering
    \scriptsize
    \begin{minipage}[t]{0.49\textwidth}
        \centering
        \captionof{table}{Complexity analysis of model size and runtime on the sequence \textit{soldier}}
        \vspace{-2mm}
        \setlength{\tabcolsep}{1.5pt}
        \renewcommand{\arraystretch}{1.8}
        \resizebox{\textwidth}{!}{%
        \begin{tabular}{c|c|c|c|c}
        \toprule
         & \textbf{YOGA \cite{refer26}} & \textbf{DeepPCC \cite{refer27}} & \textbf{JPEG Pleno VM \cite{refer29}} & \textbf{Proposed} \\ 
        \midrule
        Model size & $169.5~\text{MB}$ & $124.3~\text{MB}$ & $64.8~\text{MB}$ & $44~\text{MB}$ \\ 
        \midrule
        Enc/Dec & $-$ & $-$ & $50.1 / 37.4~\text{s}$ & $0.77 / 1.17~\text{s}$  \\
        \bottomrule
        \end{tabular}}
        \vspace{-2.5mm}
        \label{table:soldier}
    \end{minipage}
    \hfill
    \begin{minipage}[t]{0.49\textwidth}
        \centering
        \scriptsize  
        \captionof{table}{Ablation study of channel Flip scanning in the Mamba block, and entropy model (MEM)}
        \vspace{-2mm}
        \setlength{\tabcolsep}{1pt}
        \renewcommand{\arraystretch}{1.35}  
        \resizebox{\textwidth}{!}{%
        \begin{tabular}{c|c|c|c|ccccc|cc}
        \toprule
        \textbf{Method} & \textbf{Model size} & \textbf{Channel Flip} & \textbf{Channel Flip} & \multicolumn{5}{c|}{$\text{BD-BR}~(\%)$} & \textbf{Enc} & \textbf{Dec} \\
        \cmidrule(lr){5-9} 
        & $\text{MB}$ & in Mamba Block & in MEM & D1 & Y & $1\text{-}\text{PCQM}$ & GraphSim & PQI & $\text{s}$ & $\text{s}$ \\
        \midrule
        MEGA-PCC  & $44.0~\text{MB}$  & $\checkmark$ & $\checkmark$  & 0 & 0 & 0 & 0  & 0 & 1.08 & 1.63 \\ 
        Method1 & $42.7~\text{MB}$  & $\times$ & $\checkmark$  & 4.0 & 1.3 & 3.9 & 3.5 & 3.1 & 1.00 & 1.48  \\ 
        Method2  & $43.7~\text{MB}$  & $\checkmark$ & $\times$ & 3.6 & 6.4  & 9.5 & 8.9 & 10.1 & 0.92 & 1.43 \\
        Method3  & $42.5~\text{MB}$  & $\times$ & $\times$  & 6.7 & 9.8 & 12.9 & 11.3  & 12.2 & 0.82 & 1.31\\ 
        \bottomrule
        \end{tabular}}
        \vspace{-2.5mm}
        \label{table:single-decoder}
    \end{minipage}
\end{table*}


\begin{table*}[h]
    \scriptsize
    \begin{minipage}[t]{0.49\textwidth}
       \centering
        \captionof{table}{Ablation study on group size in Mamba block}
        \vspace{-3mm}
        \label{table:left}
        \setlength{\tabcolsep}{5pt}
        \renewcommand{\arraystretch}{0.8}
        \resizebox{\linewidth}{!}{%
        \begin{tabular}{c|ccccc}
            \toprule
            \multicolumn{1}{c|}{\textbf{Mamba}} & \multicolumn{5}{c}{\textbf{BD-BR (\%)}} \\\cmidrule(lr){2-6}
            \textbf{block size} & D1 & Y & 1-PCQM & GraphSim & PQI  \\
            \midrule
           \text{[2048, 1024, 512]} & -81.1 & -40.2 & -43.8 & -33.4 & -57.7  \\
            \text{[1024, 512, 256]}& -80.4 & -43.9 & -49.6 & -38.7 & -60.5  \\
            \text{[512, 256, 128]} & -81.5 & -41.5 & -45.0 & -34.6 & -59.0 \\
            \bottomrule
        \end{tabular}}
        \vspace{-5mm}
         \label{table:extract}
    \end{minipage}
    \hfill
    \centering
    \scriptsize
    \begin{minipage}[t]{0.49\textwidth}
        \centering
        \captionof{table}{Ablation study on group size in MEM}
        \vspace{-3mm}
        \label{table:left}
        \setlength{\tabcolsep}{5pt}
        \renewcommand{\arraystretch}{1}
        \resizebox{\linewidth}{!}{%
        \begin{tabular}{c|ccccc|cc}
            \toprule
            \multicolumn{1}{c|}{\textbf{MEM}} & \multicolumn{5}{c|}{\textbf{BD-BR (\%)}} & \textbf{Enc} & \textbf{Dec} \\\cmidrule(lr){2-6} 
             \textbf{group size}& D1 & Y & 1-PCQM & GraphSim & PQI & (s) & (s) \\
            \midrule
            16  & -80.4 & -43.9 & -49.6 & -38.7 & -60.5 & 1.08 & 1.63 \\
             24 & -80.9 & -44.0 & -48.6 & -38.9 & -60.4 & 1.25 & 1.97 \\
             32 & -80.8 & -44.9 & -49.6 & -39.1 & -60.8 & 1.53 & 2.28 \\
            \bottomrule
        \end{tabular}}
        \vspace{-5mm}
        \label{table:memgroup}
    \end{minipage}
    
\end{table*}
\subsection{Rate-Distortion Performance Comparison}

To comprehensively assess the effectiveness of our proposed MEGA-PCC method, we compare it against the classical standards G-PCCv23 \cite{refer2} and V-PCCv22 \cite{refer1}, as well as recent learning-based joint compression methods, including YOGA \cite{refer26}, DeepPCC \cite{refer27}, Unicorn \cite{refer31, refer32}, and JPEG Pleno VM4.1 \cite{refer29} (see Table~\ref{fig:Taxonomy}). The evaluation leverages multiple quality metrics for performance assessment. Table~\ref{table:BDBR} summarizes Bjøntegaard Delta Rate (BD-Rate) results for D1-PSNR, Y-PSNR, and 1-PCQM, with all values reported relative to G-PCCv23. Figure~\ref{fig:RDcurve} shows the corresponding rate-distortion (R-D) curves evaluated with 1-PCQM. Additional results for D2-PSNR, YUV-PSNR, GraphSIM, and PQI are available in the supplementary.

Experimental results demonstrate that MEGA-PCC substantially outperforms G-PCC across all metrics. Specifically, it achieves an average BD-Rate saving of 80.4\% in D1-PSNR, 43.6\% in Y-PSNR, and 49.6\% in 1-PCQM over the G-PCC baseline.  Compared to V-PCC, MEGA-PCC delivers a clear advantage in high bitrate scenarios, as shown in Figure~\ref{fig:RDcurve}. Moreover, it maintains competitive performance with JPEG Pleno, YOGA, and DeepPCC, while operating with significantly lower computational cost. From Table~\ref{fig:Taxonomy}, these baseline methods involve the computationally expensive recoloring. In particular, for 11-bit sequences, MEGA-PCC outperforms both YOGA and DeepPCC, which perform worse than G-PCC in this setting. This result highlights MEGA-PCC’s stronger generalization capability to different bit-depths.

Although MEGA-PCC does not surpass Unicorn~\cite{refer31, refer32}, the results of which are unverifiable because the code is unavailable, in rate-distortion performance, its design is far more efficient. Unicorn leverages a complex multi-scale entropy model and larger network architecture, resulting in higher computational demands and longer runtime. In contrast, MEGA-PCC achieves strong compression performance with a simpler and faster model, as detailed in the supplementary materials.

\subsection{Complexity Analysis}

Table~\ref{table:soldier} presents the complexity analysis for the sequence soldier, including model size and runtime. Among the learning-based joint compression methods, only JPEG Pleno VM 4.1~\cite{refer29} provides publicly available inference code, making it the sole baseline for a complete runtime comparison. For a fair runtime comparison, we tested our model only on the soldier sequence because the other sequences had been used by JPEG Pleno VM for training. Compared to MEGA-PCC, JPEG Pleno VM 4.1~\cite{refer29} not only has a larger model size but also a significantly longer processing time, partly due to certain modules being implemented on CPU. In particular, the recoloring step alone takes 2.18 seconds and is included in the reported encoding time.  In contrast, MEGA-PCC features a lightweight design with substantially lower encoding and decoding times. In addition, MEGA-PCC eliminates the need for recoloring and model matching between geometry and attribute streams during inference. These steps are typically required in other pipelines.

\subsection{Ablation Study}

The ablation study is conducted on all test sequences across the full bit-rate range to ensure comprehensive evaluation.
\noindent\textbf{Channel Flip Scanning in Mamba.}
While most point cloud compression methods focus on spatial dependencies, channel-wise correlations are often underexplored. MEGA-PCC adopts a tri-directional scanning strategy, incorporating channel-wise Flip SSM in both the encoder-decoder and the Mamba-based entropy model (MEM). To analyze the contribution of each component, we evaluate three configurations: Method1, with channel flip scanning applied only in MEM; Method2, with flip scanning applied only in the Mamba block; Method3, without flip scanning in either module. As shown in Table~\ref{table:all}, using channel flip scanning in both modules (MEGA-PCC) achieves the best rate-distortion performance, with a slight runtime increase that is justified by the gain in compression efficiency.
\vspace{0.2cm} 

\noindent\textbf{Group size in Mamba block.}
We investigate the impact of different group sizes \(\mathbf{M}\) = [$\text{M}_{1}$, $\text{M}_{2}$, $\text{M}_{3}$] in Mamba block, which define the number of tokens processed per group at each scale. As shown in Table~\ref{table:extract}, we evaluate three configurations. Among them, the configuration [1024, 512, 256]
yields the best overall rate-distortion performance across various metrics. This result indicates that a balanced block size provides a better trade-off between local detail capture and global context modeling, making it the most effective setting for our feature extraction module.
\vspace{0.1cm} 

\noindent\textbf{Group size in MEM.}
The Mamba entropy model divides serialized feature sequences into groups of size $J$, processed independently to ensure parallelism. Table~\ref{table:memgroup}  reports the effect of varying $J$. Larger group sizes improve compression performance slightly by expanding the temporal receptive field, but also increase runtime. Based on a trade-off between performance and computational cost, we select $J$=16 as the optimal setting, offering strong compression gains with minimal added complexity.

\noindent\textbf{Dual vs. Single Decoder.} 
{To validate the effectiveness of our dual-decoder design, we implemented a single-decoder variant with a similar Mamba backbone. The dual-decoder achieves an average bitrate reduction of 40.9\%, 52.1\%, and 44.7\%, compared to the single-decoder in terms of D1-PSNR, Y-PSNR, and 1-PCQM, respectively, highlighting its effectiveness in improving attribute quality and overall compression performance. See Supplementary Sec. G for more results.

\vspace{0.05cm}


\vspace{-2mm}
\section{Conclusion}
\vspace{-2mm}


In this work, we introduced MEGA-PCC, a unified and efficient framework for point cloud compression that jointly addresses geometry and attribute coding within a single architecture. By sharing representations in the encoder and decoding geometry ahead of attribute reconstruction, our approach promotes tighter integration and improved efficiency. The incorporation of sparse convolutions and a multi-directional Mamba backbone, along with the proposed Mamba-based Entropy Model (MEM), enables MEGA-PCC to capture rich spatial and channel-wise dependencies while maintaining linear-time decoding and accurate entropy modeling. Extensive experiments validate its effectiveness in achieving strong rate-distortion performance with low complexity. Future directions include among others enhancing the context modeling for even greater compression gains, extending the framework to handle dynamic point clouds with temporal coherence, exploring broader dynamic datasets (e.g.~large-scale sparse LiDAR data), and variable-rate coding.

{
    \small
    \bibliographystyle{ieeenat_fullname}
    \bibliography{main}
}


\end{document}